# Anisotropic swelling of elastomers filled with aligned 2D materials

Mufeng Liu[1], Suhao Li[1], Ian A. Kinloch[1], Robert J. Young[1] and Dimitrios G. Papageorgiou[1,2*]

[1] National Graphene Institute and Department of Materials, School of Natural Sciences, The University of Manchester, Oxford Road, Manchester M13 9PL, UK
[2] School of Engineering and Materials Science, Queen Mary University of London, Mile End Road, London E1 4NS, UK

E-mail: d.papageorgiou@qmul.ac.uk

**Abstract**

A comprehensive study has been undertaken on the dimensional swelling of graphene-reinforced elastomers in liquids. Anisotropic swelling was observed for samples reinforced with graphene nanoplatelets (GNPs), induced by the in-plane orientation of the GNPs. Upon the addition of the GNPs, the diameter swelling ratio of the nanocomposites was significantly reduced, whereas the thickness swelling ratio increased and was even greater than that of the unfilled elastomers. The swelling phenomenon has been analyzed in terms of a modification of the Flory-Rhener theory. The newly-derived equations proposed herein, can accurately predict the dependence of dimensional swelling (diameter and thickness) on volume swelling, independent of the type of elastomer and solvent. The anisotropic swelling of the samples was also studied in combination with the evaluation of the tensile properties of the filled elastomers. A novel theory that enables the assessment of volume swelling for GNP-reinforced elastomers, based on the filler geometry and volume fraction has been developed. It was found that the swelling of rubber nanocomposites induces a biaxial constraint from the graphene flakes.

## 1. Introduction

Crosslinked rubbers swell when they come in contact with liquids. The swelling of rubbers impacts negatively their mechanical properties and ultimately makes the materials lose their serviceability. It is therefore crucial to improve the liquid barrier properties for rubber-based materials to control their swelling behaviour. In the past, carbon black has been used extensively for this purpose and detailed studies were undertaken by Kraus [1]. The theory proposed by Kraus has been shown to be applicable to carbon blacks and other types of spherical fillers [2-4].

The reduced swelling ratio of carbon black-reinforced elastomer composites as the result of an improved stiffness modulus is central to Kraus' theory [1]. Carbon black can be approximated to a spherical filler that reinforces elastomers isotropically. However, when elastomers are reinforced by asymmetric fillers with unidirectional orientation such as aligned rods and platelets, the mechanical reinforcement is anisotropic [5, 6]. The anisotropy of modulus leads to anisotropic swelling when the nanocomposites are immersed in liquids [5, 6]. In this case, Kraus' equation is not applicable, because the reinforcement from asymmetric fillers is no longer uniform and leads to complexity in the swelling process.

Coran *et al.* studied the swelling behaviour of rubbers reinforced by unidirectionally-aligned fibers [5]. The authors reported that the rubbers swelled less in the axial direction of fibers (parallel) but more in the perpendicular direction to fibers (transverse). Additionally, Nardinocchi *et al.* undertook simulation of the anisotropic swelling of fiber-filled thin gels [6]. Both studies [5, 6] reported that the reason for anisotropic swelling was the dissimilar modulus in the different principal axes in the specimens. It is therefore not surprising that anisotropic swelling phenomena have not only been observed in polymer composites but also in a heterogeneous gel [7, 8] and polymer gel [9] systems, where the modulus of the material differs across the principal axes. Interestingly, both the aforementioned experiments and simulations revealed that the dimension(s) of filled elastomers swelled to a greater extent than the unfilled elastomers in the direction(s) perpendicular to filler orientation [5, 6]. It is surprising therefore that no explicit theory has been proposed to explain and quantify the increased swelling transverse to the direction of filler orientation and account for why the transverse swelling increases with increasing filler content.

The work of Treloar in 1950 reported that, counterintuitively, the application of tensile strains (uniaxial and biaxial) to rubber samples led to an increased volume swelling ratio in solvents [10]. Treloar successfully quantified the dependence of volume swelling ratio (solvent uptake) on the applied strains (either uniaxial or biaxial), on the basis of Flory and Rhener's statistical study on rubber elasticity [11, 12]. Such a potentially useful theory has, however, yet to be applied for the study of the swelling behaviour of rubber composites.

Since graphene was first isolated and identified in Manchester in 2004 [13], one obvious application was its use as a reinforcement in polymer-based nanocomposites [14, 15]. Moreover, due to its large aspect ratio and highly asymmetric geometry, graphene can be used to tailor the properties of matrix materials in different directions [16-23]. Such systems have been studied in detail over recent years and graphene nanocomposites with elastomer-based matrices have been found to show particular promise [24-29]. Since graphene and other 2D materials display a plate-like morphology it is found that they can be oriented in different directions in a polymer matrix [30] and this is known to affect the mechanical properties of the nanocomposites [31].

In this present study, three different types of elastomers (natural rubber (NR), nitrile butadiene rubber (NBR) and a thermoplastic elastomer (TPE)) were compounded with graphene nanoplatelets (GNPs) of different lateral sizes of 5 μm, 15 μm and 25 μm as quoted by the supplier (designated as M5, M15 and M25, respectively) in order to investigate their swelling behaviour in different solvents. In particular it has been found that as a result of the processing method, most of the GNPs were aligned in the plane of the elastomer sheets which led to anisotropic swelling. Some elastomer samples filled with carbon black were also tested in order to compare the swelling characteristics for different filler morphologies. A theoretical analysis has also been carried out based on the Flory-Rhener theory and the analysis of strain-induced swelling by Treloar, to fully understand the anisotropic swelling of the GNP-filled elastomers and relate this to the mechanical properties of the filled elastomers. Finally, a new equation has been derived, that enables us to assess the reinforcing efficiency of GNPs in the volume swelling of elastomers based on both the aspect ratio and volume fraction of the filler.

## 2. Results and Discussion

### 2.1 Materials characterization

The actual weight fractions of the fillers in the different elastomers were confirmed by TGA and the volume fractions are listed in **Table S4 (Supporting Information)**. The microstructure of the filled materials was characterized by observing the scanning electron microscopy images of cryo-fractured sections of the specimen, as shown in **Figure 1.** It can be seen from **Figure 1 (a-c)** that the GNPs appear edge-on protruding from the fracture surfaces with a high degreee of alignment in the plane of the elastomer sheets as a consequence of the compression moulding technique employed. On the other hand, the NR sample filled with carbon black shows carbon black clusters of submicron sizes within a relatively smooth surface of the cross-section of the sample, as seen from **Figure 1 (d)**. Typical fracture surfaces for other filler (GNP and CB) loadings and sizes of the nanocomposites showing similar characteristics are also presented in **Figures S1 and S2**.

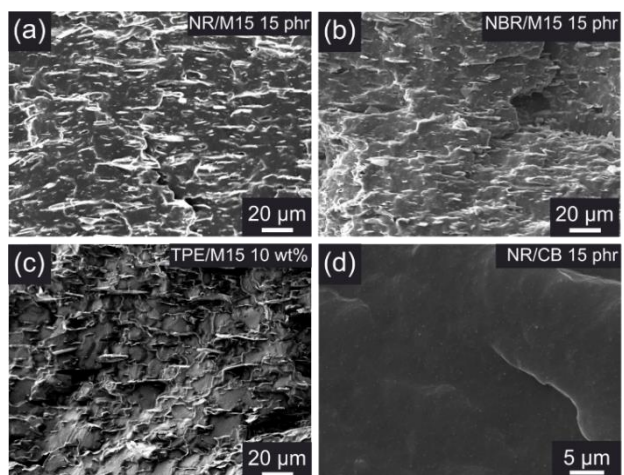

**Figure 1.** The microstructure of GNP filled nanocomposites samples: (a) NR/M15 15 phr; (b) NBR/M15 15 phr; (c) TPE/M15 10 wt% and (d) NR/CB 15 phr.

### 2.2 Dimensional Swelling

The mass uptake is defined as $M(t) = \frac{W(t)-W(0)}{W(0)}$, where $W(0)$ and $W(t)$ are the weight of dry samples and the weight measured at time $t$, respectively. The equilibrium of the swelling processes was indicated by the plateau of the curves of $M(t)$ against time. The mass uptake data for four representative samples is shown in **Figure 2** and the mass uptake data for all samples tested is shown in **Figure S3**. The addition of the fillers led to a reduction in both the gravimetric diffusion of the solvents, that was indicated by the changes in the initial slope before the saturation plateau, and the obvious differences in the mass uptake of the solvents at saturation.

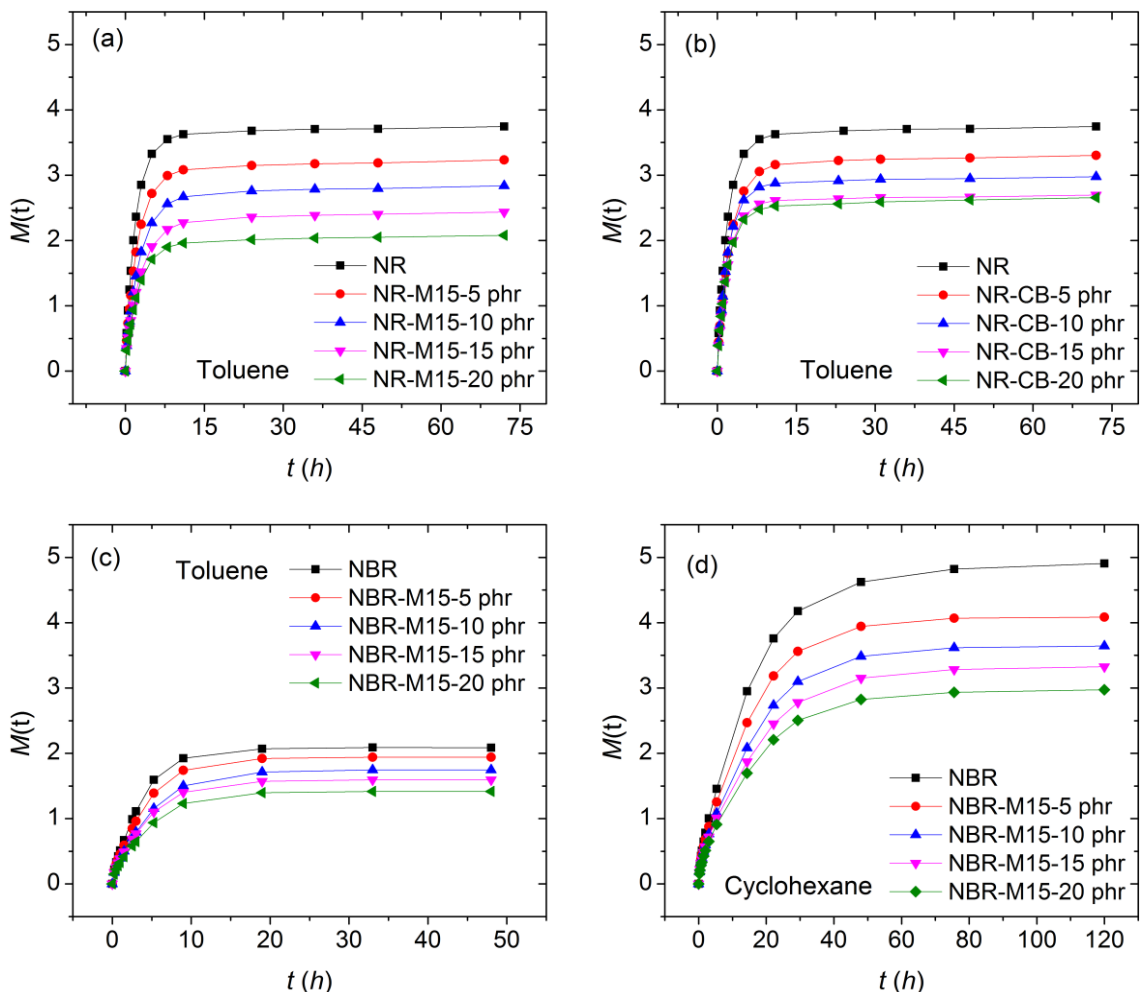

**Figure 2.** Mass uptake of toluene against time of the measurements for the nanocomposites samples: (a) NR-M15, (b) NR-CB, (c) NBR-M15; mass uptake of cyclohexane against time of (d) NBR-M15. The mass uptake is defined as $M(t)=\frac{W(t)-W(0)}{W(0)}$, where $W(0)$ and $W(t)$ are the weight of dry samples and the weight of the samples measured at time $t$, respectively.

The speed of the diffusion of the solvents was reduced due to the complex tortuous path that the solvent molecules had to diffuse with increasing filler contents. The mass uptake at saturation of swelling, on the other hand, is related to the volume absorption of the solvent by the elastomers. Therefore, after confirmation of the saturation point of swelling for the samples under study, we focus on the dimensional (volumetric) swelling, shown in **Figure 3**. From **Figure 2**, it can be seen that the total mass uptake, related to the volume swelling at saturation, for different elastomers in different solvents can vary significantly. For example, NR absorbed more toluene than NBR, comparing Figure 2(a) to (c), and NBR absorbed more cyclohexane than toluene, comparing Figure 2(c) and (d). This observation is related to the fact that different elastomers/solvents can have different solubility parameters, according to the theory proposed by Flory and Rhener [12]. The dimensionless parameter $\chi$ from Flory and Rhener's theory is related to the solubility parameters of both an elastomer and a solvent, to reflect this phenomenon.

The measurements of the sample dimensions were carried out both upon the unswollen samples and also after the swelling process reached equilibrium. **Figure 3** shows the swelling ratios of the volume, diameter and thickness at the equilibrium plotted against the volume fraction of filler, for the four respresentative samples: NR reinforced by **(a)** M15 GNPs and **(b)** carbon black, both immersed in toluene, and **(c)** NBR reinforced by M15 GNPs swollen in toluene and **(d)** cyclohexane. The swelling ratios for all the samples studied are shown in **Figure S4**.

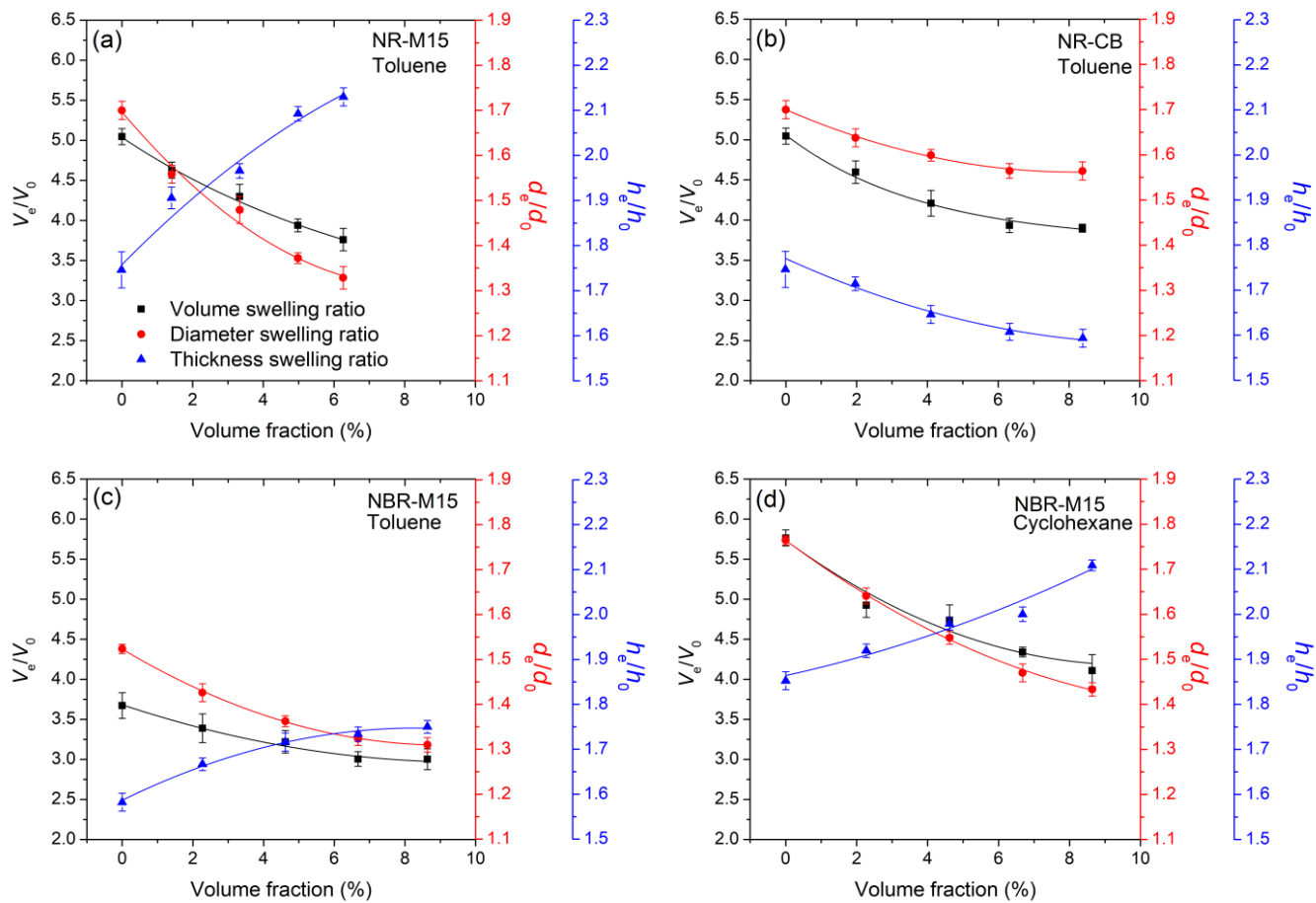

**Figure 3.** Swelling ratio at the equilibrium of the volume ($V_e/V_0$), diameter ($d_e/d_0$) and thickness ($h_e/h_0$) against the volume fraction of the filler of the nanocomposites for (a) NR-M15-GNP swollen in toluene, (b) NR-CB swollen in toluene, (c) NBR-M15-GNP swollen in toluene and (d) NBR-M15-GNP swollen in cyclohexane. The lines are a guide to the eye.

It is apparent that the addition of both GNPs and carbon black (CB) into the elastomers leads to a reduction of the volume swelling ratio ($V_e/V_0$) and therefore improved the liquid barrier properties of the elastomers. If we compare **Figure 3 (a)** with (**b**), it can be seen that the GNPs reinforced natural rubber more efficiently than carbon black. This is attributed to the higher aspect ratio of the GNPs that formed a larger interfacial area per volume (specific interfacial area) than the carbon black nanoparticles in the nanocomposites. When the elastomers started swelling, the GNPs provided a higher restraining force than CB to reduce the swelling ratio [1]. However, the GNP-reinforced elastomers swelled anisotropically, while the CB-filled materials swelled more or less isotropically. More specifically, as shown in **Figure 3 (a)**, both the diameter ($d_e/d_0$) and thickness ($h_e/h_0$) swelling ratios of the unfilled natural rubber were similar (~1.70) at the equilibrium. With increasing GNP loading, the diameter (in-plane) swelling ratio decreased, whereas the thickness (out-of-plane) swelling ratio increased. Such a phenomenon is believed to take place due to the in-plane orientation of the nanoplatelets that originated from the compression moulding process [24]. The microstructure of both thermoset rubbers and thermoplastic elastomers shown in the SEM graphs in **Figure 1** and **Figure S1 and 2** clearly demonstrates the preferred in-plane orientation of the flakes. In addition, the high degree of orientation of the GNPs in rubber/GNP nanocomposites produced with the exact same procedure, has been confirmed in our previous work [24], using advanced techniques including polarised Raman spectroscopy and X-ray computed tomography (CT) scans. The in-plane-aligned nanoplatelets carried the stress from the in-plane direction of the nanocomposites and hence restrained the diameter swelling efficiently. It is interesting to note, however, that the GNP-filled rubbers displayed higher thickness swelling ratio than the unfilled rubber, similarly to what has already been reported for fibre-reinforced rubbers [5, 6]. Similar anisotropic swelling behaviour can be observed in **Figure 3 (c)** and (**d**) for GNP-filled NBR, swollen in different solvents (toluene and cyclohexane). From **Figure 3** it can also be concluded that the swelling ratio also depends on the types of both matrix and solvent. **Figures 3 (a)** and **(c)** show the swelling behaviours in toluene of both NR and NBR reinforced by M15 GNP. The volume swelling ratio of NBR-M15 samples is clearly lower than NR-M15 samples. When the solvent is changed to cyclohexane, the NBR-M15 samples exhibited greater volume swelling ratio than in toluene, which can be realised by comparing **Figures 3 (c)** and **(d)**. This phenomenon will be explained in the next section on the basis of the statistical mechanics approach proposed by Flory and Rhener [11, 12].

It should also be noted that the lateral size of the GNPs plays a role in improving the liquid barrier properties of the nanocomposites. It can be seen in **Figure S4** that both NR and TPE nanocomposites filled with the M25 GNPs showed lower volume ($V_e/V_0$) and diameter ($d_e/d_0$) swelling ratios than the samples filled with M5 GNPs. This is because the larger flakes possess higher aspect ratio, which is beneficial for effective stress transfer, and therefore they are able to restrain the swelling of the elastomers more efficiently [32].

*2.3 Theoretical analysis of anisotropic swelling*

We have undertaken a theoretical analysis of the anisotropic swelling of the filled elastomers based on the Flory-Rhener and Treloar's studies of rubber elasticity [10-12]. The equations that have been derived enable us to determine and predict the anisotropic swelling of elastomers reinforced by oriented-GNPs and to understand how nanoplatelets reinforce elastomers biaxially. The principle of the theory is shown in **Figure 4**, where the dimensional changes of the swollen rubber are linked to the change of solvent uptake. Assuming that the volumes of rubber and solvent in the swollen material are additive, the volume fraction of the solvent in the swollen rubber is given by,

$$\phi_1 = \frac{n_1 v_1}{n_1 v_1 + n_2 v_2} \quad (1)$$

and the volume fraction of the rubber in the swollen sample is,

$$\phi_2 = \frac{n_2 v_2}{n_1 v_1 + n_2 v_2} \quad (2)$$

where $n_1$ and $n_2$ are the respective numbers of moles of the solvent and the elastomer in the swollen gel at equilibrium; $v_1$ and $v_2$ are the molar volumes of the solvent and the elastomer, respectively [12, 33]; These volume fractions can also be related to the change in dimensions of the disc-shaped sample during swelling and in particular the volume fraction of rubber in the swollen material at equilibrium is given by,

$$\phi_2 = \frac{\frac{\pi}{4} d_0^2 h_0}{\frac{\pi}{4} d_e^2 h_e} \quad (3)$$

where $d_0$ and $d_e$ are the diameters of the disc at unswollen and equilibrium swollen state, while $h_0$ and $h_e$ are the thickness of the disc at unswollen and equilibrium swollen state. These parameters can be seen in **Figure 4 (a)**. The volume swelling ratio is also equal to the reciprocal of the volume fraction of rubber in the swollen material since,

$$\frac{\frac{\pi}{4} d_e^2 h_e}{\frac{\pi}{4} d_0^2 h_0} = (d_e/d_0)^2 (h_e/h_0) = 1/\phi_2 \quad (4)$$

The addition of GNPs leads to a reduction of the volume of solvent uptake of $\delta n_1 v_1$ as the result of biaxial (in-plane) stress transfer from the matrix to the nanoplatelets (**Figure 4b**). It is assumed that the overall force contributed by the filler gives rise to a constraint of the elastomers during swelling that leads to the reduction of the number of moles of solvent uptake by $\delta n_1$. If we consider that the reduction of solvent uptake is caused by a combination of the overall in-plane and out-of-plane forces, then the in-plane force will reduce the solvent uptake by $\delta n_{1\parallel} v_1$ and the out-of-plane force reduces the solvent uptake by $\delta n_{1\perp} v_1$. As illustrated in **Figure 4 (c)**, the out-of-plane stress ($\sigma_\perp$) from the GNPs contributes an overall constraining force on the area of the circular plane ($\frac{\pi}{4} d_e^2$) at the equilibrium of $\sigma_\perp \frac{\pi}{4} d_e^2$, resulting in a reduction of the thickness at equilibrium of $\delta h_e$, and consequently a reduction of the number of moles of solvent uptake by $\delta n_{1\perp}$. Hence, the work done by the out-of-plane force is given by,

$$\delta W_\perp = \sigma_\perp \frac{\pi}{4} d_e^2 \cdot \delta h_e \quad (5)$$

Based on the relationship between $\delta h_e$ and $\delta n_{1\perp}$ shown in equation (4), equation (5) can be expressed as,

$$\delta W_\perp = \sigma_\perp \cdot \delta n_{1\perp} v_1 \quad (6)$$

**Figure 4 (c)** shows the constraining effect of in-plane-aligned nanoplatelets on the swelling in the out-of-plane direction, where the swelling process involves only a small degree of deformation of the elastomer. In this case, the change in the Helmholtz free energy $\Delta A_\perp$ (for this small reduction of thickness of $\delta h_e$) is considered to be equal to the work done by the force, $\delta W_\perp$ [33]. At constant temperature and pressure, considering that the net volume (liquid and elastomer) change is zero during swelling, the work done ($\delta W_\perp$) gives a change ($\delta$) in the Gibbs free energy of the swelling process ($\Delta G_\perp$). Thus,

$$\delta \Delta G_\perp = \Delta A_\perp = \delta W_\perp = \sigma_\perp \delta n_{1\perp} v_1 \quad (7)$$

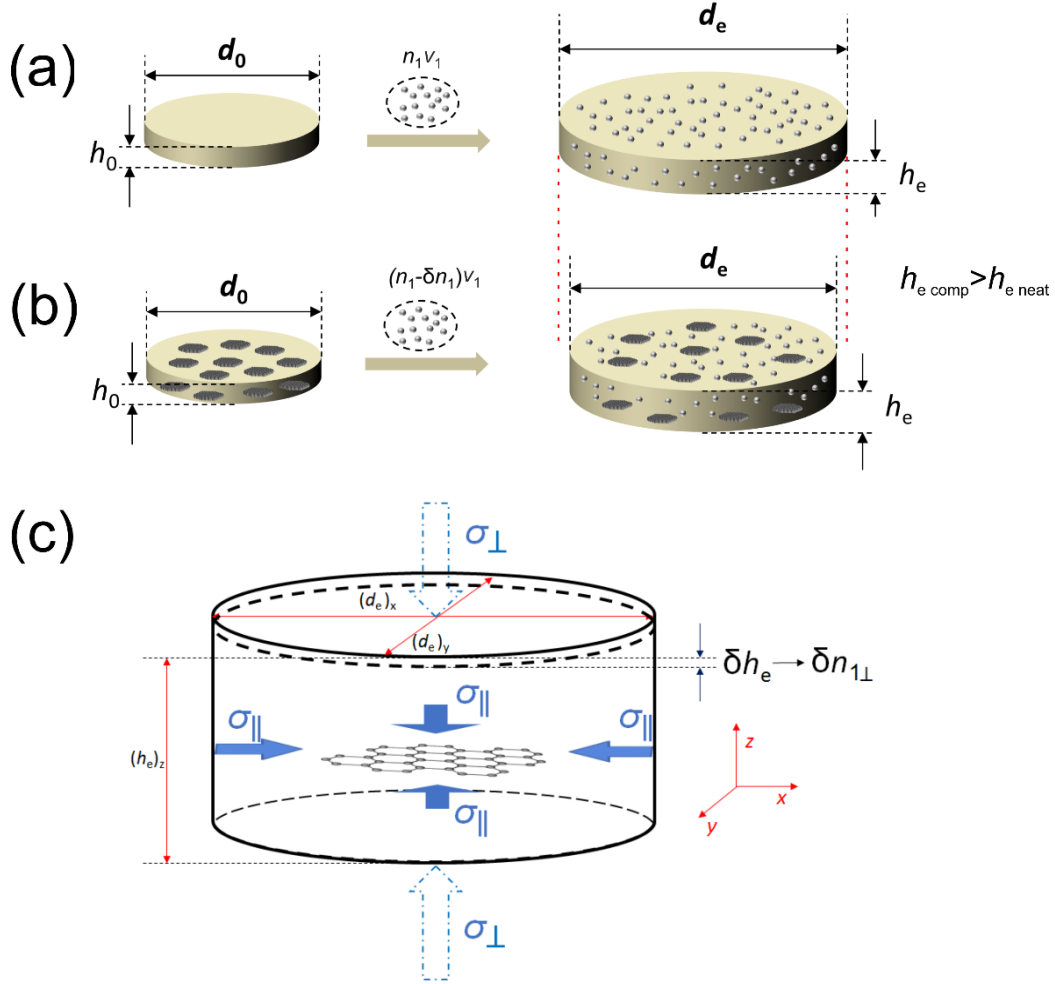

**Figure 4**. (a) Schematic diagram of the swelling of an unfilled elastomer and (b) an elastomer filled with biaxially-aligned 2D materials, where $d_0$ and $d_e$ are the diameters at the unswollen state and at the equilibrium of swelling, respectively; $n_1$ is the number of moles of the absorbed solvent molecules, $v_1$ is the molar volume of the solvent, $\delta n_1$ is the change of moles of the solvent absorbed by the elastomer after addition of GNPs. (c) Schematic diagram of the swollen state of the GNP-filled elastomer, where $\sigma_\perp$ and $\sigma_\parallel$ are the stresses contributed by the GNPs at the cross-plane and in-plane directions, respectively. The principal axes $x$, $y$, $z$ have been defined in the figure. The diameters of the discs, $(d_e)_x$ and $(d_e)_y$, and the thickness of the disc, $(h_e)_z$, were used order to express the 3-dimensional deformation of swelling process, where $(d_e)_x$ and $(d_e)_y$ should be equal.

In the process of swelling for a crosslinked polymer network, the Gibbs free energy change ($\Delta G$) should be a combination of the free energy of dilution ($\Delta G_m$) and the free energy of elastic deformation ($\Delta G_{def}$) of the materials from the unswollen network to the equilibrium swollen state, such that $\Delta G = \Delta G_m + \Delta G_{def}$ [12, 33, 34]. In terms of partial differentials, equation (7) can then be written as,

$$\left(\frac{\partial \Delta G}{\partial n_1}\right)_{(d_e)_x (d_e)_y} = \sigma_\perp v_1 = \frac{\partial \Delta G_m}{\partial n_1} + \left(\frac{\partial \Delta G_{def}}{\partial n_1}\right)_{(d_e)_x (d_e)_y} \tag{8}$$

where $\partial \Delta G_m / \partial n_1$ is free energy of mixing (dilution) from uncrosslinked unswollen state to the crosslinked equilibrium state [34]. The subscripts $(d_e)_x$ and $(d_e)_y$ represent the in-plane diameters in different principal axes as illustrated in **Figure 4 (c)**. The Flory-Huggins theory gives [24],

$$\frac{\partial \Delta G_m}{\partial n_1} = RT[\ln(1-\phi_2) + \phi_2 + \chi \phi_2^2] \tag{9}$$

where R is the gas constant, $T$ is the thermodynamic temperature, and $\chi$ is a dimensionless parameter which is dependent upon the polymer-solvent interaction. The term $\Delta G_{def}$ refers to the free energy of elastic deformation of the materials from unswollen network to the equilibrium swollen state. According to the theory of rubber elasticity for the deformation of Gaussian network [12], for a disc-shape sample with a volume of $(\frac{\pi}{4} d_0^2 h_0)$ we have for equilibrium swelling,

$$\Delta G_{\text{def}} = \frac{\pi}{4} d_0^2 h_0 \frac{\rho RT}{2M_c} \left[ \left(\frac{d_e}{d_0}\right)_x^2 + \left(\frac{d_e}{d_0}\right)_y^2 + \left(\frac{h_e}{h_0}\right)_z^2 - 3 \right] \quad (10)$$

where $\rho$ is the density of the elastomer and $M_c$ is the molar mass of the network chains between crosslinks. The subscripts $x$, $y$ and $z$ represent the principal axes as in **Figure 4 (c)**, where $(d_e/d_0)_x$ should be equal to $(d_e/d_0)_y$. Combining equation (10) with equations (2 and 3), the partial differential equation (8) can be solved to give:

$$\left(\frac{\partial \Delta G_{\text{def}}}{\partial n_1}\right)_{(d_e)_x (d_e)_y} = \left(\frac{\partial \Delta G_{\text{def}}}{\partial h_e}\right)\left(\frac{\partial h_e}{\partial n_1}\right)_{(d_e)_x (d_e)_y}$$
$$= \frac{\rho RT}{M_c} \frac{h_e}{h_0} \frac{v_1}{(d_e/d_0)^2} = \frac{\rho RT}{M_c} v_1 \phi_2 \left(\frac{h_e}{h_0}\right)^2 \quad (11)$$

Subsequently, by combining equations (8), (9) and (11), the stress from the GNPs in the out-of-plane direction ($\sigma_\perp$) is given by,

$$\sigma_\perp = \frac{RT}{v_1}[\ln(1-\phi_2) + \phi_2 + \chi \phi_2^2] + \frac{\rho RT}{M_c} \phi_2 \left(\frac{h_e}{h_0}\right)^2 \quad (12)$$

and if we substitute equation (4) into (12)

$$\sigma_\perp = \frac{RT}{v_1}[\ln(1-\phi_2) + \phi_2 + \chi \phi_2^2] + \frac{\rho RT}{M_c \phi_2} \left(\frac{d_e}{d_0}\right)^{-4} \quad (13)$$

As suggested from the results in **Figure 3**, the swelling in the out-of-plane direction (thickness) was not restrained by GNPs but actually became higher than that of the unfilled elastomer. Hence, we can assume that the nanoplatelets are able to reinforce only in the in-plane directions of the nanocomposite discs during the swelling process and the elastomer is not subjected to any mechanical constraint from the GNPs in the out-of-plane direction. The stress given by the flakes in the out-of-plane direction ($\sigma_\perp$, shown in **Figure 4c**) is therefore assumed to be 0. Consequently, we have modified the original Flory-Rhener equations for the case of in-plane aligned GNP reinforced elastomers to give,

$$[\ln(1-\phi_2) + \phi_2 + \chi \phi_2^2] + \frac{\rho v_1}{M_c}\left(\frac{h_e}{h_0}\right)^2 \phi_2 = 0 \quad (14)$$

or expanding the logarithm term,

$$\frac{\rho v_1}{M_c} = \left(\frac{1}{2} - \chi\right)\left(\frac{h_e}{h_0}\right)^{-2} \phi_2 \quad (15)$$

and,

$$[\ln(1-\phi_2) + \phi_2 + \chi \phi_2^2] + \frac{\rho v_1}{M_c (d_e/d_0)^4 \phi_2} = 0 \quad (16)$$

or expanding the logarithm term,

$$\frac{\rho v_1}{M_c} = \left(\frac{1}{2} - \chi\right)\left(\frac{d_e}{d_0}\right)^4 \phi_2^3 \quad (17)$$

Equation (16) mirrors the dependence of biaxial strain on the volume swelling ratio for rubbers swollen in solvents under biaxial tension suggested by Treloar [10]. For unfilled elastomers with uniform structures, isotropic swelling is expected. In this case we have (based on equation (4)) $d_e/d_0 = h_e/h_0 = (1/\phi_{2\text{ neat}})^{1/3}$ and both equations (15) and (17) lead to the well-accepted Flory-Rhener theory [12, 34],

$$\rho V_1/M_c \approx \left(\frac{1}{2} - \chi\right)(1/\phi_{2\text{ neat}})^{-5/3} \quad (18)$$

where $(1/\phi_{2\text{ neat}})$ is the volume swelling ratio of a neat elastomer (as $\phi_2$ is the volume fraction of the elastomer of the swollen gel at the equilibrium of swelling).

By combining equation (18) with equations (15) and (17), we obtain the final equations for the diameter and thickness swelling ratios of elastomers reinforced with oriented 2D materials,

$$d_e/d_0 = (1/\phi_{2\text{ neat}})^{-5/12} (1/\phi_2)^{3/4} \quad (19)$$

$$h_e/h_0 = (1/\phi_{2\text{ neat}})^{5/6} (1/\phi_2)^{-1/2} \quad (20)$$

where $(1/\phi_2)$ is the overall volume swelling ratio for nanocomposites. When the material is a neat polymer, equations (19) and (20) are transformed to $d_e/d_0 = h_e/h_0 = (1/\phi_{2\text{ neat}})^{1/3}$, and the swelling of the materials obeys the equation for isotropic swelling (equation 18).

Equations (19) and (20) are able to predict the dependence of the in-plane and out-of-plane dimensional swelling ratios on the volume swelling ratio for elastomers reinforced by 2D materials, assuming that the fillers are well-oriented in-plane. The products $(d_e/d_0)$ and $(h_e/h_0)$ are the respective swelling ratios in the in-plane (diameter) and cross-plane (thickness) directions at equilibrium. It is very important to point out that the newly-derived equation (20) for the thickness swelling ratio predicts an increasing trend with decreasing volume swelling ratio, in agreement with the experimental results reported in **Figure 3 (a,c,d)**. The products $(1/\phi_{2\text{ neat}})^{-5/12}$ and $(1/\phi_{2\text{ neat}})^{5/6}$ contain the information upon the polymer-solvent interaction $\chi$, while based on equation (18), other parameters including the density of polymer $\rho$, the molar volume of solvent $v_1$ and the molar mass between crosslinks $M_c$ cancel out and so they do not affect in-plane and out-of-plane swelling.

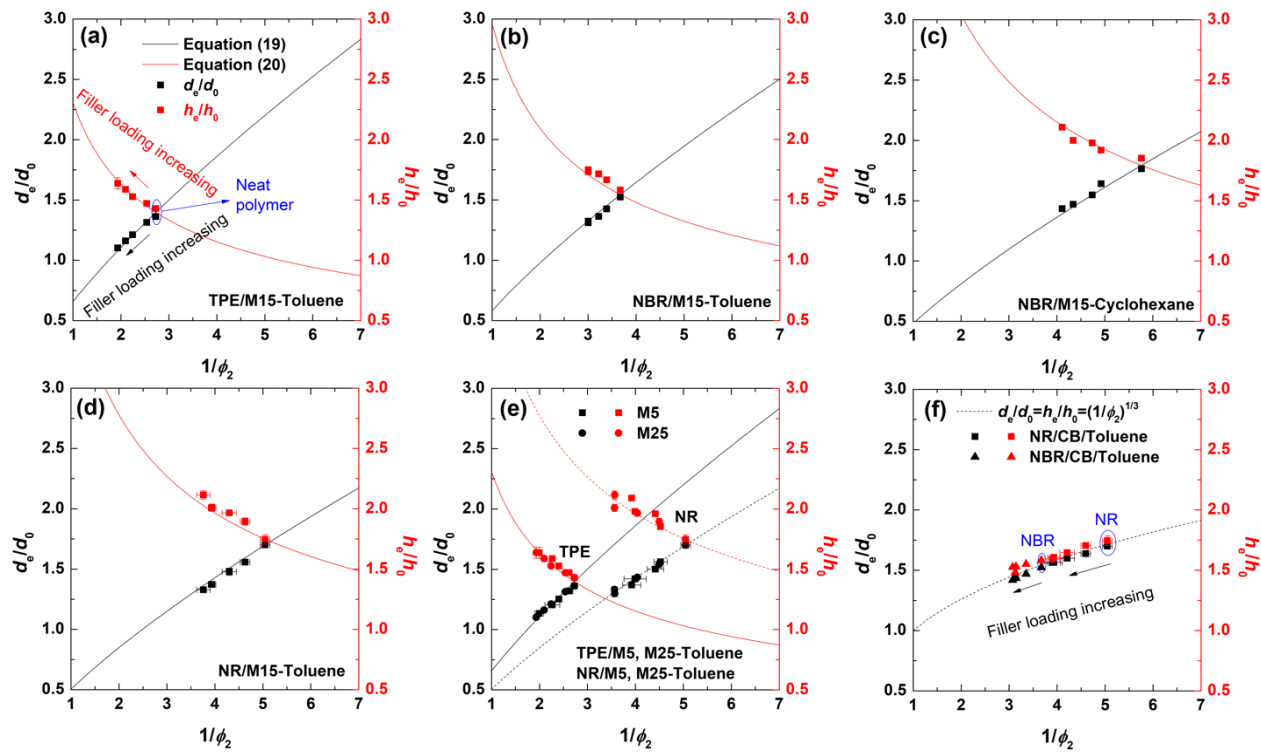

**Figure 5.** The dependence of dimensional swelling ratios ($d_e/d_0$, $h_e/h_0$) on volume swelling ratios ($1/\phi_2$) for the samples (a) TPE/M15 GNP in toluene, (b) NBR/M15 GNP in toluene, (c) NBR/M15 in cyclohexane, (d) NR/M15 in toluene, (e) TPE/M5, TPE/M25 GNP and NR/M5, NR/M25 GNP in toluene and (f) NR/CB and NBR/CB in toluene. The plotted theoretical curves in (a-e) are based on equations 19 (black) and 20 (red), in order to demonstrate the predicted dependence of diameter (in-plane) and thickness (cross-plane) swelling ratio on the volume swelling ratio ($1/\phi_2$), respectively. The theoretical curve in (f) predicts isotropic swelling, where $d_e/d_0=h_e/h_0=(1/\phi_2)^{1/3}$, for carbon black filled elastomers.

## 2.4 Application of the anisotropic swelling theory

The application of equations (19) and (20) to the experimental results for various combinations of rubber/filler/solvent are shown in **Figure 5 (a-e)**. It can be seen that the theoretical predictions demonstrate excellent consistency with the anisotropic swelling behavior of different elastomers filled with graphene nanoplatelets, while the samples were tested in 2 different solvents including toluene and cyclohexane. For the neat polymer, as highlighted (with a blue arrow) in **Figure 5 (a)**, the intersection of the two curves of equations (19) and (20) represents the theoretical dependence of $d_e/d_0$ and $h_e/h_0$ on the volume swelling ratio ($1/\phi_2$), where $d_e/d_0=h_e/h_0=(1/\phi_2)^{1/3}$. As for the experimental results, the diameter and thickness swelling ratios of the neat polymer showed a slight difference compared to the theoretical values. This can be attributed to a small degree of orientation of the chains of the amorphous polymer that was induced during compression moulding [35, 36]. With increasing filler loading, the volume and the diameter swelling ratios were reduced; however, an increase of the thickness swelling ratio was observed as shown in **Figure 5 (a)**. Almost identical swelling behaviors were revealed for all other types of elastomers filled with graphene nanoplatelets and swollen in different solvents (**Figure 5 b-e**). Regarding the CB-filled samples (NR and NBR matrices), the dimensional swelling should be isotropic, since carbon black is a spherical filler. Therefore, a theoretical curve based on the relationship: $d_e/d_0=h_e/h_0=(1/\phi_2)^{1/3}$ is plotted in **Figure 5 (f)**. The experimental results again exhibit excellent consistency with the theory.

From the analysis of the experimental results and the application of the newly-proposed equations, we have quantified the anisotropic swelling phenomenon in elastomers reinforced by aligned 2D materials, where the thickness swelling ratio increases with increasing filler loading. The physical meaning of this phenomenon is that the macromolecular network of the elastomers should be treated as incompressible, with a Poisson's ratio of ~0.5 [33]. In the process of swelling under small strains, the deformation of the rubber networks can be modeled by that of a Gaussian network (equation 10) [11, 12, 33, 34]. In this case, when a biaxial strain is applied in the two principal axes of a material, there is a consequential Poisson strain along the other principal axis.

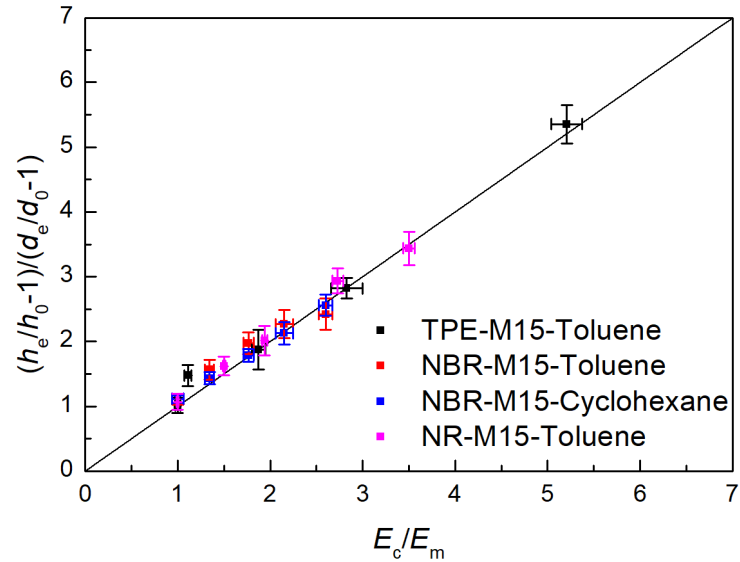

**Figure 6.** The swelling strain ratios [$(h_e/h_0-1)/(d_e/d_0-1)$] against the normalized modulus ($E_c/E_m$) obtained from tensile tests for all the elastomers filled with M15 GNPs.

For the nanocomposites in the present study, it can be understood that the graphene nanoplatelets enabled the application of a biaxial constraining strain on the elastomer matrices during swelling by in-plane stress (equivalent to lateral compression). As a consequence, this resulted in increasing strain in the out-of-plane axis. Since increased solvent uptake can be induced by increasing tensile strain [33], it can be understood that the elastomers absorbed less solvent in the directions where the GNPs are oriented (in-plane), and on the other hand, absorbed more solvent in the direction perpendicular to the filler orientation (out-of-plane) experiencing the Poisson strain.

*2.5 Reinforcing efficiency of the GNPs*

We have successfully analysed the anisotropic swelling induced by GNPs orientation. It is possible to quantify the individual parameters of the GNPs that determine the swelling ratio of the nanocomposites. In order to achieve this, we have examined comparatively the anisotropic modulus and anisotropic swelling of the nanocomposites.

The osmotic (swelling) pressure ($\Pi$) from the solvent to the elastomers is given by [33],

$$\Pi = -(RT/v_1)\ln(p/p_0) \tag{21}$$

where R is the gas constant, $T$ is the Kelvin temperature, $v_1$ is the molar volume of the pure liquid, $p$ is the vapour pressure of the liquid component in equilibrium with the mixture (swollen rubber) and $p_0$ is the saturation vapour pressure of pure liquid. The swelling pressure is dependent only upon the type of the solvent and defined as the pressure that keeps an elastomer swollen at equilibrium [33]. Hence, the swelling pressures from the solvent to the swollen elastomer should be the same for either the parallel ($\Pi_\parallel$) or perpendicular ($\Pi_\perp$) directions, relative to the GNP orientation (in-plane). For small strains in the swelling process,

$$\Pi_\parallel = \Pi_\perp = E_\parallel \varepsilon_\parallel = E_\perp \varepsilon_\perp \tag{22}$$

where $E_\parallel$ and $E_\perp$ are the uniaxial moduli of the nanocomposites parallel ($x$ or $y$) and perpendicular ($z$) to the filler orientation, respectively, and $\varepsilon_\parallel$ and $\varepsilon_\perp$ are the corresponding strains at equilibrium of swelling. Replacing the swelling strains with the swelling ratios, we have,

$$\frac{E_\parallel}{E_\perp} \approx \frac{h_e/h_0-1}{d_e/d_0-1} \tag{23}$$

The same relationship was also obtained by Coran *et al.* for rubber/carbon fibre composites under swelling [5]. The stiffness parallel to the filler orientation ($E_\parallel$) is related to the modulus values measured parallel to the rubber sheets, which can be considered as the modulus of the composites ($E_c$). The stiffness perpendicular to the filler orientation can be considered equal to the stiffness of the neat matrix ($E_m$), since it was found that the GNPs did not constrain the swelling in the out-of-plane direction. Hence, theoretically, we should have $\frac{E_c}{E_m} \approx \frac{E_\parallel}{E_\perp} \approx \frac{h_e/h_0-1}{d_e/d_0-1}$. A similar method to link the results of mechanical tests to swelling testing was also carried out by Goettler *et al.* [37]. To examine this relationship, the mechanical properties were obtained by tensile tests and the stress-strain curves are presented in **Figure S5**. The modulus values for all the samples are listed in **Table S5**.

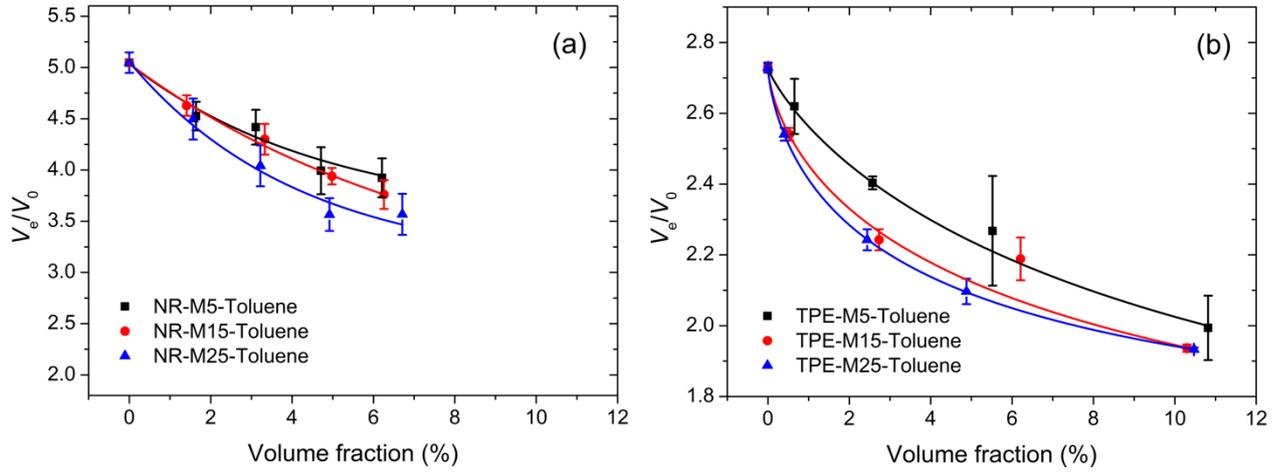

**Figure 7.** Volume swelling ratio ($V_e/V_0$) against volume fraction of the GNPs for (a) NR and (b) TPE, reinforced by M5, M15 and M25 GNPs. The increasing aspect ratio of the filler results in reducing volume swelling of the nanocomposites.

The swelling strain ratios ($\frac{h_e/h_0 - 1}{d_e/d_0 - 1}$) obtained from swelling tests are plotted in **Figure 6** against the corresponding normalized modulus ($E_c/E_m$) for representative elastomer/GNP nanocomposites at different GNP loadings. The same figure that includes all samples under study is presented in **Figure S6**. It can be seen that the datapoints lie predominantly on the line with a slope of 1, indicating a linear relationship between the values of swelling strain ratios and the normalized moduli. Therefore, we can assume that equation (23) is applicable for the study of elastomer/GNP nanocomposite samples.

The dimensional swelling ratios ($d_e/d_0$, $h_e/h_0$) can be substituted by equations (19) and (20). Hence, we can develop a theoretical relationship between the volume swelling ratio and the modulus of the nanocomposites ($E_c/E_m$),

$$\frac{(1/\phi_{2\,\text{neat}})^{5/6}(1/\phi_2)^{-1/2}-1}{(1/\phi_{2\,\text{neat}})^{-5/12}(1/\phi_2)^{3/4}-1} = \frac{E_c}{E_m} \quad (24)$$

Based on our previous research, the normalized modulus ($E_c/E_m$) of an elastomer reinforced by 2D materials is dependent upon the aspect ratio ($s$), the orientation factor ($\eta_o$) and the filler volume fraction ($\phi_f$), while it is independent of the stiffness of the filler [26, 32]. For the swelling properties, since the filler particles are well aligned, the orientation factor ($\eta_o$) of the filler can be set equal to 1, in accordance with the experimental results in Section 3.2 and theoretical analysis in Section 3.3 and 3.4. Therefore,

$$\frac{(1/\phi_{2\,\text{neat}})^{5/6}(1/\phi_2)^{-1/2}-1}{(1/\phi_{2\,\text{neat}})^{-5/12}(1/\phi_2)^{3/4}-1} = 1 - \phi_f + 0.056 s^2 \phi_f^2 \quad (25)$$

Equation (25) provides a relationship between the volume swelling ratio ($1/\phi_2$) and the filler parameters (aspect ratio and volume fraction). Quite importantly, it is clear that with increasing filler aspect ratio (M25>M15>M5) and increasing volume fraction, the swelling ratio ($1/\phi_2$) should be reduced, which is in agreement with the results shown in **Figure 7.** The size effect of different GNPs upon the volume swelling presented in **Figure 7** clearly shows a decreasing trend with increasing filler aspect ratio but this does not seem remarkably obvious. In a previous work we performed measurements on the size of different M-grade GNPs [32], which revealed that the difference in the lateral size between the different grades of the fillers is not as large as advertised (~5, 6 and 7 μm for M5, M15 and M25). Thus, the reinforcing efficiency which is dependent on the effective aspect ratio of the fillers within the nanocomposites, is not altered drastically [26,29] from M5 to M15 to M25. It is believed that the increased lateral dimensions of a filler exert an impact on the swelling behaviour by increasing its effective aspect ratio in the nanocomposites. However, obvious differences in the properties of the nanocomposites can be clearly observed only when the effective aspect ratio between different GNPs is significantly different. It is also interesting that equation (25) shows that the swelling behavior depends only upon the volume fraction and aspect ratio of the filler ($s$), but not upon its Young's modulus. This is because 2D fillers such as GNPs are effectively infinitely stiff compared with the elastomer matrix [32].

## 3. Conclusions

In this work, a number of elastomers were filled with different graphene nanoplatelets and carbon blacks to study the swelling behaviour of the nanocomposites in solvents. Both GNPs and CB were able to restrain the volume swelling ratio of the materials. It was found that the restraining efficiency of GNPs was higher than CB in volume swelling, as a result of the high aspect ratio of GNPs that created higher interfacial area per volume. It has been demonstrated that the compression moulding process clearly contributed to the in-plane orientation of the GNPs in the elastomer sheets, leading to anisotropic swelling of the nanocomposites. The swelling measurements revealed that the GNPs only reinforced the materials in the in-plane directions while no enhancement was found in the cross-plane direction. Moreover, the dimensional swelling ratio in the cross-plane direction became even greater with increasing GNP filler loading. Elastomers reinforced by carbon blacks, however, swelled isotropically due to the symmetric (spherical) shape of the filler.

A simple and effective theory has been developed for the comprehensive study of anisotropic swelling for elastomer nanocomposites reinforced by in-plane-oriented graphene nanoplatelets. The newly-derived equations are able to accurately predict anisotropic dimensional swelling in both principal axes parallel and perpendicular to the filler orientation, independent of the type of elastomers, GNPs or solvents. The theory of anisotropic swelling can also be combined with the mechanical properties of the nanocomposites measured by tensile tests, in order to assess the reinforcing efficiency of the GNPs to the elastomers, in comparison to their swelling behaviours. The use of micromechanical theories enabled us to relate our observations to measurable physical parameters of the fillers and a correlation between anisotropic swelling and mechanical properties was successfully established. Through the comprehensive experimental and theoretical analysis presented in this study, we can conclude that the use of 2D materials (such as graphene) can enable the tailoring of the swelling characteristics of an elastomeric matrix. This can have very important implications in highly demanding elastomer-based applications such as seals operating in liquid environments, rapid gas decompression seals and o-rings and gaskets for the chemical, petroleum, automotive and aerospace industries.

## 4. Experimental Methods

### 4.1 Materials and preparation

A range of different elastomers were selected for this study that include natural rubber (NR), nitrile butadiene rubber (NBR) and a thermoplastic elastomer (TPE). The grade of natural rubber (NR) used was SMR CV60 (Standard Malaysian Rubber, Mooney-Viscosity ML (1+4, 100 °C) of 60), which was purchased from Astlett Rubber Inc., Oakville, Ontario, Canada and used as received. The nitrile butadiene rubber (NBR) was Nipol® 1052J, supplied by Clwyd Compounders Ltd and used as received. The thermoplastic elastomer, Alcryn® 2265 UT (Unfilled Translucent), which is based on a partially crosslinked chlorinated olefin interpolymer alloy, was purchased from A. Schulman, Inc, and used as received.

Graphene nanoplatelets (xGNP, Grade-M particles, XG Sciences, Lansing USA) with nominal lateral sizes of 5 μm, 15 μm and 25 μm as quoted by the supplier (designated as M5, M15 and M25, respectively) were used for compounding with the NR, NBR and TPE. The thicknesses of all the flakes were quoted by the manufacturer to be in the range 6 to 8 nm (i.e. around 20 layers of graphene). The same batches of the GNPs were fully characterised in our previous study [32]. The actual lateral size of the nanoplatelets used herein, is smaller than the advertised values. The average lateral sizes of M5, M15 and M25 GNPs were measured to be ~5 μm, ~6 μm and ~7 μm, respectively.

High abrasion furnace (HAF) N330 carbon black (CB) supplied by the Berwin Polymer Processing Group, Duckinfield, UK, was employed to mix with NR and NBR. Moreover, all the additives involved in some of the rubber processing, zinc oxide, stearic acid, TMTD (tetramethylthiuram disulfide), CBS (n-cyclohexyl-2-benzothiazole sulfenamide) accelerator and sulfur were of analytical grade and used as received. Detailed formulations are listed in the Tables S1 and S2. The solvents used in the analysis of swelling behavior of the elastomer nanocomposites, toluene (anhydrous, 99.8%), cyclohexane (anhydrous, 99.5%) and acetone (anhydrous, 99.9%) were purchased from Sigma-Aldrich Co. Ltd. and used as received. The compounding of NR and NBR was carried out in a two-roll mill at room temperature. The loadings of 5, 10, 15 and 20 phr (parts by weight per hundred parts of rubber) for M5, M15 and M25 xGNPs and N330 CB were incorporated into NR, while only M15 xGNPs and N330 were mixed with NBR. The melt mixing of the TPE with 3 types of xGNPs was carried out in a Thermo Fisher HAAKE Rheomix internal mixer. The mixing took place at 165 °C and 50 rpm for 5 minutes. The GNP loadings (M5, M15 and M25) in the TPE nanocomposites were 1%, 5%, 10%, and 20% by weight (and not phr since no other additives were used with this matrix, see Table S3).

The compounds were subsequently cut into small pieces and hot pressed in a metal mould into sheets (~2 mm thick) in a Collin Platen Press (Platen Press P 300 P/M). The vulcanization proceeded at a temperature of 160 °C for 10 minutes under a hydraulic pressure of 30 bar for NR and NBR. For TPE, the moulding took place in the same equipment at 185 °C for 10 minutes. All the moulded elastomer sheets (~2 mm thick) were then stamped into disc-shaped samples with a diameter of around 25 mm for swelling tests and into dumbell samples for tensile tests.

### 4.2 Characterisation

The actual loadings of the fillers in the nanocomposites were determined by thermogravimetric analysis (TGA) using a TA Q500 TGA instrument. The samples were heated from room temperature up to 600 °C under a 50 mL/min flow of $N_2$ at 10°C /min.

The microstructure of the cryo-fractured elastomer nanocomposites was examined using scanning electron microscopy (SEM). The images were acquired using a high-resolution XL30 Field Emission Gun Scanning Electron Microscope (FEGSEM) operated at 6 kV.

For the swelling studies, the samples were immersed in ~50 mL of solvent. The uptake of solvent was monitored by weighing the samples at intervals following immersion until their weight became constant and equilibrium was established. The dimensions of the samples were also measured using a vernier caliper in both the unswollen state and in the equilibrium swollen state. Different sets of elastomers and nanocomposites were tested in different solvents as listed in **Table 1.**

**Table 1.** The nanocomposites and solvents used for swelling measurements.

| Matrix | Filler | Loadings | Solvent |
|---|---|---|---|
| NR | GNP (M5,15, 25)/CB | 5, 10, 15, 20 phr | Toluene |
| NBR | GNP (M15)/CB | 5, 10, 15, 20 phr | Toluene, Cyclohexane |
| TPE | GNP (M5,15, 25) | 1, 5, 10, 20 wt% | Toluene |

Tensile testing was undertaken using an Instron 4301 machine with a load cell of 5 kN, for all the samples. At least 5 specimen were tested for each sample. The NR and NBR samples were tested under a tensile rate of 500 mm·min$^{-1}$ in accordance with ASTM 412 standard, while the TPE samples were tested under a tensile rate of 50 mm·min$^{-1}$ in accordance with ASTM 638.

## Acknowledgements

This project has received funding from the European Union's Horizon 2020 research and innovation programme under grant agreement No 785219. Ian A. Kinloch also acknowledges the Royal Academy of Engineering and Morgan Advanced Materials. All research data supporting this publication are available within this publication.